\documentclass[11pt]{article}

\newcommand{\system}{\textsc{AI Watchman}}

\usepackage[final]{acl}

\usepackage{times}
\usepackage{latexsym}

\usepackage[T1]{fontenc}

\usepackage[utf8]{inputenc}

\usepackage{microtype}

\usepackage{inconsolata}

\usepackage{graphicx}

\title{Accounting for Stochasticity in Studies of Large Language Model Refusal}
\author{
\textbf{Emma Lurie\textsuperscript{1}},
\textbf{Stephanie T. Wang\textsuperscript{1}},
\textbf{Sorelle A. Friedler\textsuperscript{2}},
\textbf{Danaé Metaxa\textsuperscript{1}} 
\\
\textsuperscript{1}University of Pennsylvania
\textsuperscript{2}Haverford College \\
\texttt{\{ewlurie, stephtw, metaxa\}@seas.upenn.edu}\\
\texttt{sorelle@cs.haverford.edu}
}
\begin{document}
\maketitle
\begin{abstract}
 We present preliminary empirical evidence that single-observation queries are insufficient for evaluations of LLM refusal behaviors. Using a longitudinal auditing system, we issued identical prompts 100 times each across four dates to GPT-4.1 for two socially salient topics across 20 Wikipedia sources. Refusal outcomes were consistent with a stable Bernoulli process, yet 20\% of sources fell within a decision-boundary region where a single query is largely uninformative. Reliable quantification of refusals required between 15 and 25 repeated queries, well above the single-observation standard common in existing evaluations.\end{abstract}

\section{Introduction}

Large language models (LLMs) are regularly found to create violent, hateful, or otherwise concerning text~\cite{grok2025pbs, chatgpt2025guardian}, and AI companies developing LLMs regularly take steps to reduce the prevalence of these and other undesired outputs~\cite{ahmad2025openai, markov2023holistic}. Content moderation for LLMs can include safeguards built into the model itself or the use of separate automated filters~\cite{markov2023holistic}; regardless of the technical mechanism, the goals and targets of such moderation are determined by company policies shaped by the legal and political landscape, societal norms, and corporate values~\cite{gillespie2018custodians, klonick2017new}. This paper focuses on evaluating contexts in which models abstain from answering some or all of a user's query — what model developers term \emph{refusal}~\cite{openai2024gpt4osystemcard, yuan2025hard}.

As LLMs increasingly mediate access to information about social issues for a large public audience, AI model developers' content moderation policies can limit access to information and shape public discourse based on opaque company policies.


LLMs are known to behave stochastically~\cite{bender2021dangers}, yet the implications of this stochasticity for evaluating content moderation remain underexplored. In this paper, we present empirical evidence that refusal behavior varies substantially across repeated identical prompts, and that the degree of variance differs systematically by topic and by individual source. Using a longitudinal auditing system for LLM content moderation~\cite{metaxa2021auditing}, we issued identical prompts 100 times each across four separate dates to GPT-4.1 for two socially salient topics that we observed inconsistent refusal behavior on: Abortion and Israel Global Image. Refusal outcomes across all twenty sources were consistent with a stable, per-source Bernoulli process, yet 20\% of sources fell within the decision-boundary region ($0.3 < \hat{p} < 0.7$) where a single query carries little information about the true refusal probability. We further find that reliable source-level rankings require between 15 and 25 repeated queries depending on topic, well above the single-pass standard common in existing audits.

These findings have direct implications for evaluation methodology. Single-pass benchmarks and model cards that report refusal behavior without repeated sampling may mischaracterize a model's moderation policy, and may conflate stochastic variation with genuine policy changes over time. 

\section{Related Work}
\subsection{LLM Abstention and Refusal}
Model refusals have been studied in contexts including safety, where a response may cause harm or conflict with ethical standards; knowledge gaps, where a query is ambiguous, incomplete, or falls outside the model's knowledge; and model uncertainty, where the model lacks sufficient confidence in the correctness of its response~\cite{wen_know_2025, brahman2024the}. 
Techniques promoting safe LLM interactions include filtering and moderation layers, fine-tuning, and reinforcement learning with human feedback~\cite{markov2023holistic, bianchi2024safetytuned, dai2024safe}. Various evaluation frameworks also help measure and mitigate LLM under-moderation harms~\cite{wang_-not-answer_2024, ganguli2022redteaminglanguagemodels, mazeika_harm_2024, xie2025sorrybench, han_wildguard_2024}.  

The moderation of generative AI systems can also produce \textit{over-moderation}, with  systems incorrectly rejecting or flagging safe content. An audit of OpenAI's moderation endpoint found evidence of over-moderation of television violence relative to normative expectations based on age ratings~\cite{mahomed_auditing_2024}. Recent audits have found that identity-related content is overmoderated across different automated content moderation APIs, including OpenAI's moderation endpoint and Llama Guard~\cite{proebsting_identity-related_2025}. These audits have focused on the classification systems that LLM companies apply to raw model outputs before results are shown to chat interface users, rather than direct model output text. 

\subsection{Stochasticity in AI Evaluation}
Despite growing attention to the reliability of LLM evaluations, little work has examined whether refusal behavior remains consistent across repeated runs of identical prompts under the same model configuration, and across different dates. However, prior work documents instability in LLM responses and its implications for the robustness of evaluation results. LLM behavior is known to be sensitive to changes in prompt format or phrasing \cite{rottger_political_2024, elazar_measuring_2021}; small prompt modifications have been shown to flip model outputs from refusal to compliance and vice versa \cite{rottger_xstest_2024}. Beyond prompt sensitivity, LLMs can produce different outputs for the same prompt under identical settings \cite{atil-etal-2025-non}. Even configurations intended to maximize deterministic behavior do not fully eliminate this variability \cite{ouyang_empirical_2025}. For example, setting temperature to zero does not guarantee deterministic outputs in code generation \cite{ouyang_empirical_2025}. %
\section{Methods}

\subsection{AI Watchman System}
We conduct this work using \system, a longitudinal auditing system designed to publicly measure and track LLM refusals over time. \system\ operates by prompting LLMs to repeat provided text content and recording whether the model refuses to do so. Our standard evaluation pipeline runs queries automatically on a biweekly basis across OpenAI's GPT-4.1 and GPT-5 and DeepSeek's API. Our corpus of query text contains social issue topics drawn from the Pew Research Center's research areas~\cite{pewtopics}, which cover a broad range of issues of public interest. For each topic, we identified relevant Wikipedia pages as source content, chosen for their encyclopedic tone and politically neutral framing, making it less likely that refusals would be triggered by stylistic or viewpoint reasons beyond the substance of the topic itself \cite{wiki_acceptance, wiki_editorial_policy}. 

We classify model responses as refusals in two ways. \emph{Explicit refusals} occur when the content moderation system returns a direct indicator that the content is undesirable. \emph{Non-explicit refusals} occur when the model declines via its text response, for example by stating that it cannot engage with the provided content, without a formal moderation flag. Both constitute refusals in our analyses.

\begin{figure*}[t]
\centering
\includegraphics[width=\textwidth]{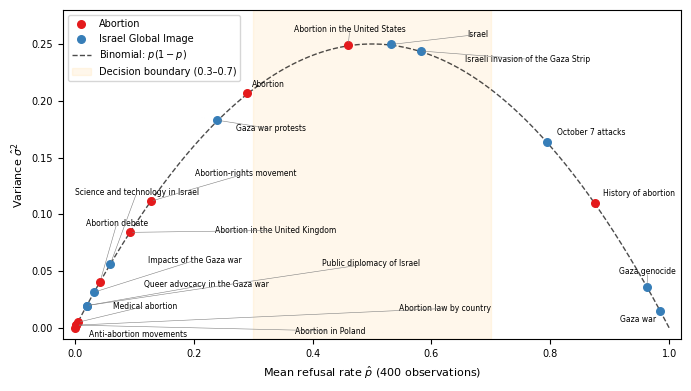}
\caption{Empirical variance vs.\ mean refusal rate for 20 Wikipedia sources (10 on Abortion, 10 on Israel Global Image) across 400 repeated queries. 
The dashed curve shows the theoretical Bernoulli variance $p(1-p)$. Points track the curve closely: refusal outcomes are consistent with an independent and identically distributed Bernoulli process. Four sources fall within the shaded decision boundary region ($0.3 < \hat{p} < 0.7$), where variance approaches its theoretical maximum
and single queries carry little information about the true refusal probability.}
\label{fig:decision_boundary}
\end{figure*}

\subsection{Stochasticity Experiment}
Given the known stochasticity of LLMs~\cite{bender2021dangers}, a single-observation pipeline may not reliably characterize a model's true refusal rate for any given topic. We designed a repeated-sampling experiment focused on GPT-4.1, which exhibited the most fluctuation in refusal rates across our longitudinal monitoring.
We selected two topics that had prior inconsistent refusal behavior in our work on \system for in-depth analysis: \textbf{Abortion} and \textbf{Israel Global Image}. For each topic, we issued identical prompts 100 times in rapid succession on four separate dates in fall 2025: 10/7, 10/20, 11/3, and 11/17. This design allows us to assess stochastic variation both within a single session (across 100 repeated prompts on the same date) and over time (across different model deployments). Each topic comprised multiple Wikipedia source pages; across both topics, we tested 20 unique sources in total.
\section{Findings}
We find two related problems with standard refusal measurement: that some sources sit at a model's decision boundary where a single observation carries near-zero signal, and that single-pass rankings are consequently unreliable.

\subsection{Decision-Boundary Sources Are Identifiable from Mean Refusal Rate}

Plotting empirical variance against mean refusal rate, we find outcomes for all twenty sources are consistent with a stationary Bernoulli process, with four sources falling within the decision-boundary region where uncertainty is highest.

For a Bernoulli random variable with success probability $p$, the variance is $\sigma^2 = p(1-p)$, maximized at $p = 0.5$. A source with a mean refusal rate near $0.5$ occupies the model's \emph{decision boundary}: the region where the model is maximally uncertain, and where a single observation provides the least information about the
underlying probability.

We compute each source's mean refusal rate $\hat{p}$ and variance $\hat{\sigma}^2$ across all 400 observations (100 queries $\times$ 4 collection
dates). Figure~\ref{fig:decision_boundary} plots $(\hat{p},\, \hat{\sigma}^2)$ for each source against the theoretical curve $p(1-p)$. Points that lie on the curve indicate sources whose refusal outcomes are
consistent with independent and identically distributed  Bernoulli draws; points above the curve would indicate overdispersion (changes over time or within-source heterogeneity).

As shown in Figure~\ref{fig:decision_boundary}, four of the twenty Wikipedia sources in this sample fall within the decision-boundary region ($0.3 < \hat{p} < 0.7$): ``Abortion'' ($\hat{p} = 0.308$, $\hat{\sigma}^2 = 0.213$), ``Abortion in the United States'' ($\hat{p} = 0.448$, $\hat{\sigma}^2 = 0.248$), ``Israel'' ($\hat{p} = 0.490$, $\hat{\sigma}^2 = 0.251$), and ``Israeli Invasion of the Gaza Strip'' ($\hat{p} = 0.545$, $\hat{\sigma}^2 = 0.249$).

All four sit at or near the theoretical maximum of $p(1-p) = 0.25$. Crucially, all points across both topics hug the binomial curve closely, indicating that refusal outcomes are well-described by a stationary Bernoulli process for each source. In this sample, the decision-boundary problem is not an artifact of changes over time, but rather a property of how the model responds to these sources.

\subsection{Single-Pass Rankings Are Insufficient}

Refusal benchmarks are routinely used to produce ordinal claims: that a model treats two subject areas differently. 
These claims implicitly rely on source rankings being stable under repeated sampling.

\begin{figure}
    \centering
    \includegraphics[width=\linewidth]{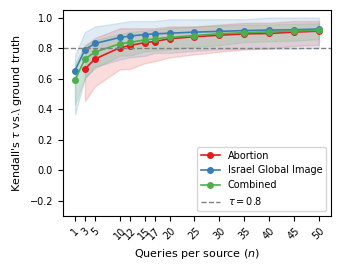}
    \caption{Mean Kendall's $\tau$ vs.\ ground-truth ranking as a function of queries per source (1,000 simulations, shaded = 95\% CI). Reliable rankings require $n \geq 15$--25 depending on topic.}    \label{fig:ranking_instability}
\end{figure}

We test ranking stability by treating the full 400-observation per-source distribution as ground truth, then simulating 1,000 trials of drawing
$n$ observations per source, ranking sources by sample mean, and computing Kendall's $\tau$ against the ground-truth ranking (see Figure~\ref{fig:ranking_instability}).

Kendall's $\tau$ measures the fraction of pairwise source orderings that agree between the sample ranking and ground truth, ranging from $-1$ (perfect reversal) to $1$ (perfect agreement), with $\tau = 0$ indicating rankings no better than random chance. We define ground truth as the average source ranking derived from all 400 observations per source, treating this as the best available estimate of each source's true underlying refusal probability.

At $n = 1$, mean $\tau$ is 0.65 for Israel Global Image and 0.59 for the combined 20-source ranking (see Figure~\ref{fig:ranking_instability}). Most Abortion sources are rarely refused, so a single observation produces near-total ties across sources and pairwise order undefined.

Defining reliability as $\tau > 0.8$ in at least 90\% of simulations, Israel Global Image sources stabilize at $n = 15$ (92.7\% of simulations), as does the combined ranking ($n = 15$, 90.2\%). Abortion sources require $n = 25$ (91.2\%). In both cases, $n = 1$ falls well below the reliability threshold. Reliable rankings require substantially more data, and the exact threshold varies by topic.

\section{Discussion}
Our results show that a single query per source is not sufficient to reliably measure LLM refusal behavior. Model refusals are stochastic, behaving like Bernoulli draws with an underlying probability of refusal, so a single observation tells us little about the underlying probability. Our ranking simulations illustrate the consequences: with only one observation per source, rankings are highly unstable, and Kendall's $\tau$ falls below reliability thresholds. Empirically our topic source rankings stabilize after 15-25 observations per source.  

More broadly, our results highlight that benchmark evaluations done with one-off prompts may be insufficient. Meaningful comparisons require repeated sampling to avoid mischaracterizing refusal rates.  

For the sources in our sample, the decision-boundary problem is likely not due to content moderation changes over time. But model updates, policy changes, or shifts in training data can still silently alter model refusal behavior. Longitudinal evaluation, which \system{} supports, remains essential for detecting such changes.  

Future work is needed to understand why particular sources fall on the decision boundary. Identifying such sources will likely require ongoing empirical monitoring across topics and further testing into the specific content in our Wikipedia sources.

These findings have implications for the design of \system{} and similar systems. First, refusal observations should be treated as probabilistic, with repeated observations collected to estimate refusal probabilities and their uncertainties. Second, systems should incorporate ranking stability checks to assess how rankings change under resampling and provide confidence measures. 

\section{Limitations}
These findings are preliminary and subject to important limitations. Our sample is small: 20 sources across two topics queried against a single model (GPT-4.1), and we cannot claim that the decision-boundary pattern, the ranking instability thresholds, or the Bernoulli characterization generalize to other topics, source types, or models.

\section{Conclusion}
Our results suggest that LLM refusal behavior is substantially more variable than single-pass evaluations imply, and that some text content is structurally more difficult to evaluate than others. Reliable audits of LLM content moderation require repeated sampling strategies.

\bibliography{custom,chi-references}



\end{document}